\documentclass[%
 preprint,
 amsmath,amssymb,
 aps,
]{revtex4-2}

\usepackage{graphicx}% Include figure files
\usepackage{dcolumn}% Align table columns on decimal point
\usepackage{bm}% bold math
\usepackage{physics}
\usepackage[normalem]{ulem}
\usepackage{xcolor}

\begin{document}

\preprint{APS/123-QED}

\title{Cold molecular ions via autoionization below the dissociation limit}% Force line breaks with \\
%\thanks{A footnote to the article title}%

\author{Sascha Schaller, Johannes Seifert, Giacomo Valtolina, Andr\'e Fielicke, Boris G. Sartakov, and Gerard Meijer} 
\email{meijer@fhi-berlin.mpg.de}

\affiliation{%
Fritz-Haber-Institut der Max-Planck-Gesellschaft, Faradayweg 4-6, 14195 Berlin, Germany\\
}%

\date{\today}% It is always \today, today,
             %  but any date may be explicitly specified

\begin{abstract}
Several diatomic transition metal oxides, rare-earth metal oxides and fluorides have the unusual property that their bond dissociation energy is larger than their ionization energy. In these molecules, bound levels above the ionization energy can be populated via strong, resonant transitions from the ground state. The only relevant decay channel of these levels is autoionization; predissociation is energetically not possible and radiative decay is many orders of magnitude slower. Starting from translationally cold neutral molecules, translationally cold molecular ions can thus be produced with very high efficiency. By populating bound levels just above the ionization energy, internally cold molecular ions, exclusively occupying  the lowest rotational level, are produced. This is experimentally shown here for the dysprosium monoxide molecule, DyO, for which the lowest bond dissociation energy is determined to be 0.0831(6) eV above the ionization energy.
\end{abstract}

%\keywords{Suggested keywords}%Use showkeys class option if keyword
                              %display desired
\maketitle

%\tableofcontents

%\section{\label{sec:level1}First-level heading:\protect\\ The line
%break was forced \lowercase{via} \textbackslash\textbackslash}

The process in which an electronically excited atom A$^*$ interacts with a ground state atom B to form the atomic ion B$^+$ and a de-excited atom A plus a free electron has been reported upon by Penning almost hundred years ago \cite{Penning1927}. This process can compete with the formation of a bound molecular ion AB$^+$ plus a free electron, known as associative ionization \cite{Arango2006}. Associative ionization can also occur when two electronically excited atoms interact \cite{Gould1988}. A less considered process is that of associative ionization of ground state atoms, i.e. A + B $\rightarrow$ AB$^+$ + e$^-$ \cite{Oppenheimer1977}. At low collision energy, the latter process is only possible when the bond dissociation energy ($D_0$) of the molecule AB is larger than its ionization energy ($IE$). For the vast majority of diatomic molecules, $IE$ is larger than $D_0$ and associative ionization cannot occur. However, there are exceptions, and the cations of several transition metal oxides and rare-earth metal oxides as well as of some fluorides can be produced from the ground state atoms via associative ionization \cite{Petrie1994,schofield2006}. These cations can be produced in yet another unique way, namely via autoionization of the neutral molecules from bound levels between $IE$ and $D_0$, adding to the various processes to generate cold molecular ions \cite{Deiß2024}.

The specific energy level structure of some diatomic transition metal oxides and rare-earth metal oxides, fluorides and their ions, together with the relativistic nature (high-$Z$) of the metal atom, makes them ideal candidates for a variety of fundamental physics studies \cite{DeMille2024}. We outline here that laser-preparation of these strongly bound neutral molecules in highly excited levels, that is, preparing these molecules with an energy above their $IE$ but below $D_0$, can be exploited to produce cold molecular ions via autoionization. Different from other schemes in which autoionization of Rydberg states \cite{Loh2011} or threshold photoionization \cite{Zhang2023} is used to produce ions, there is no competition with predissociation; this channel is closed in these molecules under these conditions. As autoionization will be many orders of magnitude faster than radiative decay, all molecules that are laser-prepared in these levels will therefore turn into a cation. This production of ions only involves resonant electronic transitions between bound levels in the neutral molecule and is therefore highly efficient. In particular, this scheme can be used to produce the cations exclusively in their lowest rotational state. We here demonstrate this on the dysprosium monoxide (DyO) molecule but note that these advantages generally hold for all neutral molecules with an $IE$ below $D_0$. Based on the literature values for $IE$ (6.395 eV; \cite{Barker2012}) and $D_0$ (6.72 eV; \cite{Lau1989}) of ThF, we expect this scheme to work for the preparation of cold ThF$^+$ cations, for instance, the species to be used in one of the next generation experiments in search of the electron's electric dipole moment \cite{Zhou2019}. For TaO$^+$, the other proposed molecular ion for searches of physics beyond the standard model \cite{Fleig2017}, the best known literature value for $D_0$ (8.70 eV; \cite{Luo2007}) is also above $IE$ (8.61 eV; \cite{Dyke1987}).

To accurately determine $IE$ when this is below $D_0$ is nontrivial as there will be many excited electronic states in the region around the ionization threshold, obscuring a clear ionization onset. An accurate spectroscopic determination of $D_0$ can be equally complicated when this is in the ionization continuum. Dissociation limits can normally be recognized from the orders-of-magnitude change in lifetime when going from bound levels to dissociating levels \cite{Sorensen2020}. When the bound levels can autoionize, however, this will occur on a similar timescale as dissociation and it is not {\it a~priori} clear how to identify $D_0$. 

In the present study we accurately determine both $IE$ and $D_0$ of the DyO molecule. From recent measurements it had been concluded that $D_0$ of DyO is 0.33 $\pm$ 0.02 eV {\it below} its $IE$ \cite{Ghiassee2023}.
We show that $D_0$ of DyO is actually 0.0831 $\pm$ 0.0006 eV {\it above} its $IE$, and that there are many bound, non-Rydberg levels of the neutral molecule in the region of $IE$. Ground state DyO$^+$ cations can be produced via autoionization from these laser-prepared, highly excited levels that are below $D_0$ in DyO.

We have performed laser ionization studies on a pulsed, jet-cooled beam of DyO molecules. Details of the experimental setup are given elsewhere \cite{Popa2024}. Several of the low-lying electronic states of DyO have been well characterized \cite{Linton1986}. Relevant for the present study is that in these states the sum $\Omega$ of the projections of the total electron spin $S$ and of the total electron angular momentum $L$ on the internuclear axis is well defined. The ground state of DyO has $\Omega$=8 and is referred to as the $X8$ state. The lowest rotational level in $X8$ has a total angular momentum of $J$=$\Omega$=8 \cite{Linton1986}. The electronically excited states of DyO are labelled according to their energy in cm$^{-1}$/1000, put in square brackets, followed by the value of $\Omega$, when known. The $X8$ state is closest to a $^7H_8$ state, i.e. with $S$=3 and $L$=5. The ground state of the DyO$^+$ cation is $^6H_{7.5}$ \cite{Ghiassee2023}.

In our experiments, we use the accurately known rotational transitions in the $[17.1]7$ $\leftarrow$ $X8$ band of DyO for the initial excitation \cite{Linton1986}. This band is known to be a $v'=0 \leftarrow v''=0$ band. Dysprosium has four main isotopes ($^{161-164}$Dy) with similar abundance; the abundance of $^{160}$Dy is one order of magnitude less. We use a reflectron time-of-flight mass spectrometer and can detect and fully separate these five DyO$^+$ isotopologues. From time-delayed two-color ionization measurements, the lifetime of the $[17.1]7$ state is found as about 50 ns.

Using pulsed laser radiation with a bandwidth of 0.07 cm$^{-1}$, all DyO isotopologues are simultaneously excited on the $P(8)$ line of the $[17.1]7$ $\leftarrow$ $X8$ band. From there, resonance enhanced multi photon ionization (REMPI) is performed with a second pulsed dye laser, time-delayed by some tens of ns. When this second laser is scanned around the wavelength of the first, preparation laser, the spectra shown in Fig. 1 are obtained. The lines observed in the spectra of $^{164}$DyO, $^{163}$DyO, $^{162}$DyO and $^{160}$DyO are due to resonant excitation to levels that are some 34100 cm$^{-1}$ above the ground-state with subsequent ionization from there (Scheme I). The width of the lines is mainly determined by the bandwidth of the second dye laser. We have identified that the $[34.1]$ state has $\Omega$=8 and that the lines with an asterisk are $R(7)$ lines of the $[34.1]8$ $\leftarrow$ $[17.1]7$ band. The anomalously large isotope shifts seem to indicate that $\Delta v$ in this band is very large and that $[34.1]8$ contains about 0.75 eV of vibrational energy; similar observations have been made for HfF \cite{Loh2012}.

\begin{figure}
\includegraphics[width=8.6cm]{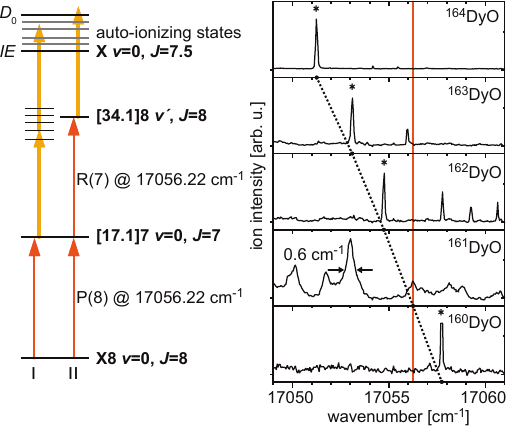}
\caption{\label{fig:1} Spectra obtained for five DyO isotopologues via (1+1)-REMPI from the laser-prepared $J$=7 level in $[17.1]7$ (Scheme I). Asterisks denote $R(7)$ lines of $[34.1]8$ $\leftarrow$ $[17.1]7$. For $^{161}$DyO resonant two-photon excitation occurs with the first laser to the $J$=8 level in $[34.1]8$. The second laser then probes rapidly decaying levels above $IE$ (Scheme II).}
\end{figure}

It is seen from the series of spectra in Fig. 1, that the frequency of the $R(7)$ line of the $[34.1]8$ $\leftarrow$ $[17.1]7$ band of $^{161}$DyO coincides with the frequency of the first, preparation laser, indicated by the red vertical line. The first laser already efficiently excites $^{161}$DyO from the $J$=8 level in $X8$ to the $J$=8 level in $[34.1]8$. The time-delayed, second dye laser then performs single-photon ionization from there (Scheme II). The lines observed for $^{161}$DyO have Lorentzian shapes and their widths are indicative of decay of the upper levels in less than 10 ps. 

In the present study, we exploit that we can prepare the $^{161}$DyO isotopologue in the $J$=8 level of $[34.1]8$ with the first, preparation laser. Using delayed ionization with the second laser, we have measured the lifetime of this level to be about 270 ns. By keeping the frequency of the first laser fixed and by using a time-delay of about 200 ns for the second laser, we exclusively obtain single photon ionization spectra from $^{161}$DyO, starting from $J$=8 in $[34.1]8$. The $^{161}$Dy isotope has a nuclear spin of 5/2 and the span of the hyperfine structure of $^{161}$DyO in $J$=8 of $X8$, $v''=1$ has been measured to be about 0.25 cm$^{-1}$; this will be similar for $J$=8 of $X8$, $v''=0$. In the $J$=7 level of $[17.1]7$ it is only about half as wide \cite{Cheng1992}. Together with the selection rules for the hyperfine components, this implies that the energy-span of the hyperfine components of $J$=8 in $[34.1]8$ that we reach is less than 0.05 cm$^{-1}$. In spite of the hyperfine splitting, we thus know the energy of the starting level in $^{161}$DyO very precisely as 34112.45 $\pm$ 0.05 cm$^{-1}$. Although this scheme is fortuitous and might appear unique, we have found and employed a similar scheme for $^{162}$DyO, in that case starting from $J$=9 in $X8$. This isotopologue has no hyperfine structure and the results obtained for $IE$ and $D_0$ are within the error bars the same as the results reported here for $^{161}$DyO. In the text, we will refer to $^{161}$Dy$^{16}$O as DyO from now on, and explicitly note it when another isotopologue is meant.

\begin{figure*}
\includegraphics[]{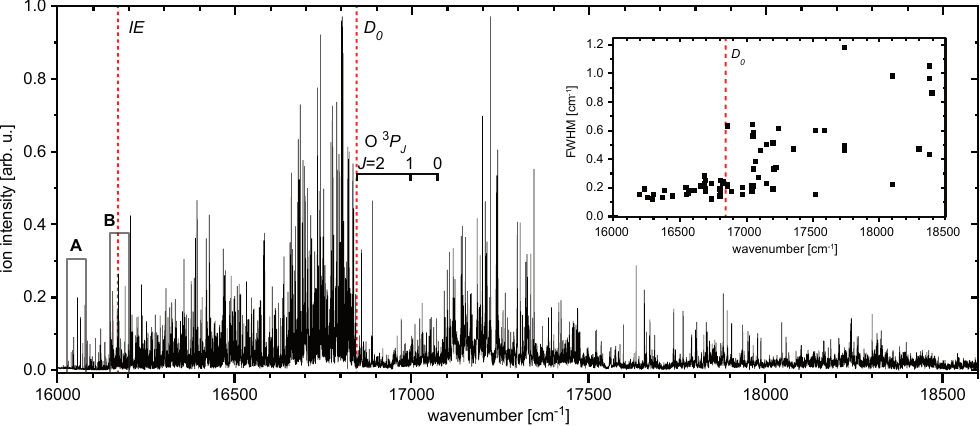}% Here is how to import EPS art
\caption{\label{fig:2} Intensity of the $^{161}$DyO$^+$ signal as a function of photon energy, when exciting from $J$=8 in $[34.1]8$ at 34112.45 cm$^{-1}$. The photon energies needed to reach the field-free $IE$ and $D_0$ are indicated by vertical dashed lines. In the inset, the full width at half maximum (FWHM) of isolated lines is shown. The spectral structure in the regions A and B is discussed in the text.}
\end{figure*}

In Fig. 2, the excitation spectrum starting from $J$=8 in $[34.1]8$ of DyO is shown. The intensity of the ion signal is recorded as a function of laser frequency over a 2500 cm$^{-1}$ spectral range in 0.05 cm$^{-1}$ steps. Laser excitation is performed in zero electric field and a pulsed extraction field $F$ of about 300 V/cm is switched on with a time-delay of 0.5 $\mu$s. The frequency needed for the third photon to reach the field-free $IE$ is indicated as such. This frequency can be extrapolated from the observed onset of ion signal for various values of $F$, assuming a square-root dependence \cite{Merkt1997}. For a hydrogenic atom, $IE$ is lowered by 6.12$\sqrt{F}$ cm$^{-1}$, with $F$ in V/cm, and we also observe a pre-factor of about 6 cm$^{-1}$ for DyO. The field-free $IE$ of DyO is determined much more precisely from multiple Rydberg series, some of which can be recognized in area A of Fig. 2. These will be discussed below, just as the region around $IE$, area B in Fig. 2. The frequency at which the third photon reaches the first dissociation limit, corresponding to ground state Dy($^5I_8$) and O($^3P_2$) atoms, can be recognized by an abrupt disappearance of the rich spectral structure around 16843 cm$^{-1}$, indicated by $D_0$. The value we thus find for $D_0$ is 50955 cm$^{-1}$, here determined to a precision of $\pm$ 5 cm$^{-1}$. The indicated O($^3P_1$) and O($^3P_0$) limits are 158 cm$^{-1}$ and 227 cm$^{-1}$ higher, respectively, whereas the next limit due to Dy($^5I_7$) and O($^3P_2$) is 4134 cm$^{-1}$ higher and no longer on this scale. The total number of electronic states of DyO converging to each of these limits is determined by $L_O$=1 of the O-atom and $L_{Dy}$=6 of the Dy-atom via (2$L_O$+1)(2$L_{Dy}$+1), which is equal to 39. Each of these electronic states will have a rich substructure with many $\Omega$-manifolds and ro-vibrational levels. Many of the 117 electronic states that converge to the Dy($^5I_8$) plus O($^3P_{J}$; $J$=0, 1, 2) limits will be attractive, some will be repulsive, and if these states have the same symmetry, avoided crossings can occur. This is the rationale for the gap in the spectrum around the energy of the three O-atom limits and this might be a defining feature to determine $D_0$ in these rare-earth metal oxides. There also is a clear increase in the widths of many lines in the spectrum, from a value that is never larger than 0.30 cm$^{-1}$ for lines below $D_0$ to values of 0.6-1.2 cm$^{-1}$ for lines above $D_0$, as shown in the inset to Fig. 2. The lines above $IE$ and below $D_0$ are slightly broadened by 0.1-0.2 cm$^{-1}$ due to autoionization, which is concluded to take place on a time-scale of tens of ps. Above $D_0$, many lines are broadened further due to dissociation, taking place on a time-scale of 5--10 ps.

In the 16000--16150 cm$^{-1}$ frequency range, several clear Rydberg series have been identified. For this, excitation is performed in zero electric field and only after 1.5 $\mu$s an electric field of several hundred V/cm is switched on to field-ionize molecules that remain in high-lying levels and to extract the ions. Due to the large electronic angular momentum in $[34.1]8$ of at least $L$=5, core-nonpenetrating Rydberg states with high angular momentum values $l$ that have no detectable quantum defect are preferentially excited. The energies $E(n,J)$ of the members of the Rydberg series with principal quantum number $n$ that converge to the rotational level $J$ in the cation can therefore be expressed as
\begin{equation}
E(n,J) = IE + B_0 [ J\cdot(J+1) - 7.5\cdot8.5] - \frac{Ry(\text{DyO})}{n^2}
\end{equation}
where $B_0$ is the rotational constant in the electronic and vibrational ground state of DyO$^+$ and where $Ry(\text{DyO})$ = 109736.977 cm$^{-1}$ is the Rydberg constant of DyO. We have observed Rydberg series converging to the lowest eight rotational levels in the cation, i.e. from $J$=7.5 to $J$=14.5. These partly overlapping series can be best recognized in the range from $n$=25--45. Fitting these Rydberg series to eq. (1) yields a value of $IE$=50284.57 $\pm$ 0.07 cm$^{-1}$ for DyO; the frequency needed for the third photon to reach $IE$ starting from $J$=8 in $[34.1]8$ is 16172.12 $\pm$ 0.03 cm$^{-1}$. The rotational constant is found as $B_0$=0.3791 $\pm$ 0.0001 cm$^{-1}$.

\begin{figure}
\includegraphics[width=8.6cm]{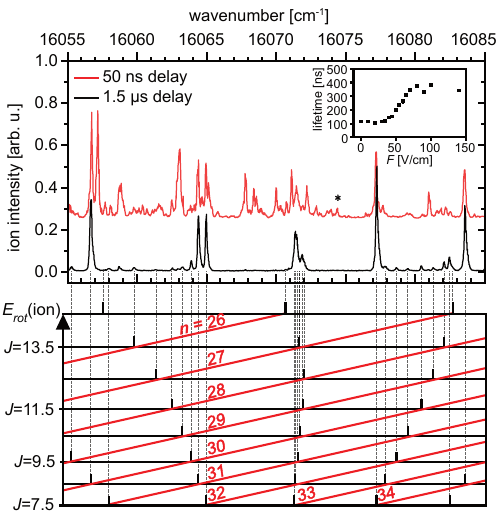}
\caption{\label{fig:3} Spectra recorded by excitation from $J$=8 in $[34.1]8$ of $^{161}$DyO to levels about 100 cm$^{-1}$ below $IE$. Excitation is in zero electric field; field-ionization and ion-extraction is 50 ns (red) or 1.5 $\mu$s (black) later. The Rydberg peaks observed in the lower spectrum are identified by their principal quantum number $n$ and by $J$ of the cation in the panel underneath. The lifetime of the level indicated by the asterisk is given in the inset as a function of an applied weak electric field, $F$.}
\end{figure} 

In Fig. 3, a 30 cm$^{-1}$ section of these Rydberg series centered about 100 cm$^{-1}$ below $IE$ is shown as the black curve. This particular region is interesting as the Rydberg peaks $E(n-i,J+i)$ with $n$=33 and $J$=7.5 all fall within 1 cm$^{-1}$ for $i$=0-6. On either side of these overlapping Rydberg peaks, there is a region of about 6 cm$^{-1}$ that is completely void of Rydberg peaks. The Rydberg peaks $E(n-i,J+i)$ with $n$=32 and $J$=7.5 can be recognized as a series of red-shifted peaks for $i$=0-6, while for $n$=34 and $J$=7.5 a series of blue-shifted peaks is seen for $i$=0-6. In the red curve, the spectrum is shown when a delay of 50 ns is used between excitation and field-ionization and extraction. This spectrum shows transitions to many other, high-lying levels in the DyO molecule. By scanning the time delay between the excitation laser and the switching on of the electric field, the lifetime of most of these extra levels is found to be in the 50--200 ns range. To distinguish these levels from the long-lived Rydberg levels, we will refer to these as the "short-lived levels". The lifetime of the short-lived level indicated by an asterisk is shown in the inset as a function of a weak, DC electric field. The lifetime of this level is seen to be around 100 ns and to remain at this value for electric fields below 50 V/cm, but to increase rather abruptly for electric fields larger than this. We attribute this to interaction of this short-lived level with the Stark-shifted components of the neighboring Rydberg levels. For the hydrogen atom in an external electric field, the high-field seeking Stark components of the $n$=34 level cross the low-field seeking Stark components of the $n$=33 level for an electric field of about 40 V/cm; this is the so-called Inglis-Teller limit. Although the level indicated with an asterisk is furthest away from any Rydberg level, it cannot escape the lifetime lengthening due to the interaction with the Rydberg levels when the electric field is very similar to this value. For all other "short-lived levels" in this spectral region the lifetimes abruptly increase at much lower values of the electric field already. 

\begin{figure}
\includegraphics[width=8.6cm]{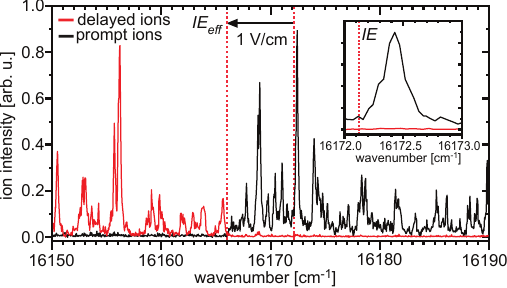}
\caption{\label{fig:4} Intensity of the $^{161}$DyO$^+$ signal as a function of photon energy when exciting from $J$=8 in $[34.1]8$ to the region around $IE$. All observed lines are to short-lived, high-lying non-Rydberg levels of the neutral DyO molecule. The region just above the field-free $IE$ is shown enlarged in the inset.}
\end{figure}

The electric field of the Inglis-Teller limit scales with the distance to $IE$ to the power -2.5, and for levels that are less than 20 cm$^{-1}$ below $IE$ an electric field of 1 V/cm is already sufficient to cause severe mixing of the "short-lived levels" with nearby Rydberg levels. This explains the observations shown in Fig. 4, when prompt ions and ions that are produced via field ionization after a certain time-delay are separately detected. Normally, the prompt ions show a steep ionization onset whereas the signal of the delayed ions appears as a somewhat broadened peak just below $IE$, caused by field-ionization of high-$n$ Rydberg levels \cite{Zhu1991}. Instead, the signals due to the delayed ions and the prompt ions shown in Fig. 4 just appear as a continuous spectrum. It is the excitation spectrum, starting from the $J$=8 level in $[34.1]8$, to the many short-lived, highly excited non-Rydberg levels of neutral DyO around $IE$, that is observed in different channels. For levels below $IE$, the weak electric field of 1 V/cm that is needed to separate the prompt and delayed ions is sufficient to mix them with Rydberg levels. This makes them long-lived and appear in the delayed, field-ionized ion channel. Levels above the effective $IE$ in 1 V/cm, which is about 6 cm$^{-1}$ below the field-free $IE$, can autoionize and appear in the prompt ion channel.  

The strongest peak in the spectrum in Fig. 4 is less than 0.4 cm$^{-1}$ above the field-free $IE$, as shown enlarged in the inset. When exciting on this peak under field-free conditions, DyO$^+$ is efficiently produced via autoionization. These ions will exclusively populate the $J$=7.5 level, as this is the only level that is energetically accessible; the $J$=8.5 level is more than 6 cm$^{-1}$ higher in energy. Excitation can also be performed in a weak electric field, tuned such that a strong transition to a level just above the effective $IE$ in that field can be used for excitation and autoionization. When an electric field of 1 V/cm is used, for instance, any of the peaks observed in the spectrum shown in Fig. 4 that lie between the dashed vertical lines can be used to produce DyO$^+$ ions that are exclusively in their lowest rotational state. 

In summary, we have accurately determined $IE$ of DyO as 6.234495(9) eV, about 0.15 eV higher than the best known value thus far \cite{Ackermann1976}. The rotational constant in the vibrational ground state of $^{161}$Dy$^{16}$O$^+$ has been found as 0.3791(1) cm$^{-1}$, corresponding to an internuclear separation of 1.748 \AA; the theoretical bond length ranges between 1.712--1.758 \AA~\cite{Ghiassee2023}. We have determined $D_0$ of DyO as 6.3176(6) eV, slightly below the literature value of 6.37 eV \cite{Luo2007}, placing $D_0$ of DyO 0.0831(6) eV above its $IE$. From the thermochemical cycle $D_0$(DyO) + $IE$(Dy) = $D_0$(DyO$^+$) + $IE$(DyO) together with the accurately known value for $IE$ of the Dy atom of 5.939064(6) eV \cite{Studer2017} we find that $D_0$ of DyO$^+$ is 6.0222(6) eV, considerably larger than recently reported \cite{Ghiassee2023}. Molecules that have their $IE$ below $D_0$, will have many bound, non-Rydberg levels in the region around $IE$. As shown here for DyO, these levels can be populated via strong transitions and enable the efficient production of state-selected, internally cold molecular ions via autoionization, their only relevant decay channel.

\begin{acknowledgments}
G.V. acknowledges support from the Alexander von Humboldt Foundation and the European Union (ERC, LIRICO 101115996).
\end{acknowledgments}

%apsrev4-2.bst 2019-01-14 (MD) hand-edited version of apsrev4-1.bst
%Control: key (0)
%Control: author (8) initials jnrlst
%Control: editor formatted (1) identically to author
%Control: production of article title (0) allowed
%Control: page (0) single
%Control: year (1) truncated
%Control: production of eprint (0) enabled
%

%\nocite{*}
%\bibliography{Schaller} % Produces the bibliography via BibTeX.

\begin{thebibliography}{26}%
\makeatletter
\providecommand \@ifxundefined [1]{%
 \@ifx{#1\undefined}
}%
\providecommand \@ifnum [1]{%
 \ifnum #1\expandafter \@firstoftwo
 \else \expandafter \@secondoftwo
 \fi
}%
\providecommand \@ifx [1]{%
 \ifx #1\expandafter \@firstoftwo
 \else \expandafter \@secondoftwo
 \fi
}%
\providecommand \natexlab [1]{#1}%
\providecommand \enquote  [1]{``#1''}%
\providecommand \bibnamefont  [1]{#1}%
\providecommand \bibfnamefont [1]{#1}%
\providecommand \citenamefont [1]{#1}%
\providecommand \href@noop [0]{\@secondoftwo}%
\providecommand \href [0]{\begingroup \@sanitize@url \@href}%
\providecommand \@href[1]{\@@startlink{#1}\@@href}%
\providecommand \@@href[1]{\endgroup#1\@@endlink}%
\providecommand \@sanitize@url [0]{\catcode `\\12\catcode `\$12\catcode
  `\&12\catcode `\#12\catcode `\^12\catcode `\_12\catcode `\%12\relax}%
\providecommand \@@startlink[1]{}%
\providecommand \@@endlink[0]{}%
\providecommand \url  [0]{\begingroup\@sanitize@url \@url }%
\providecommand \@url [1]{\endgroup\@href {#1}{\urlprefix }}%
\providecommand \urlprefix  [0]{URL }%
\providecommand \Eprint [0]{\href }%
\providecommand \doibase [0]{https://doi.org/}%
\providecommand \selectlanguage [0]{\@gobble}%
\providecommand \bibinfo  [0]{\@secondoftwo}%
\providecommand \bibfield  [0]{\@secondoftwo}%
\providecommand \translation [1]{[#1]}%
\providecommand \BibitemOpen [0]{}%
\providecommand \bibitemStop [0]{}%
\providecommand \bibitemNoStop [0]{.\EOS\space}%
\providecommand \EOS [0]{\spacefactor3000\relax}%
\providecommand \BibitemShut  [1]{\csname bibitem#1\endcsname}%
\let\auto@bib@innerbib\@empty
%</preamble>
\bibitem [{\citenamefont {Penning}(1927)}]{Penning1927}%
  \BibitemOpen
  \bibfield  {author} {\bibinfo {author} {\bibfnamefont {F.~M.}\ \bibnamefont
  {Penning}},\ }\bibfield  {title} {\bibinfo {title} {{\"U}ber {I}onisation
  durch metastabile {A}tome},\ }\href@noop {} {\bibfield  {journal} {\bibinfo
  {journal} {Naturwissenschaften}\ }\textbf {\bibinfo {volume} {15}},\ \bibinfo
  {pages} {818} (\bibinfo {year} {1927})}\BibitemShut {NoStop}%
\bibitem [{\citenamefont {Arango}\ \emph {et~al.}(2006)\citenamefont {Arango},
  \citenamefont {Shapiro},\ and\ \citenamefont {Brumer}}]{Arango2006}%
  \BibitemOpen
  \bibfield  {author} {\bibinfo {author} {\bibfnamefont {C.~A.}\ \bibnamefont
  {Arango}}, \bibinfo {author} {\bibfnamefont {M.}~\bibnamefont {Shapiro}},\
  and\ \bibinfo {author} {\bibfnamefont {P.}~\bibnamefont {Brumer}},\
  }\bibfield  {title} {\bibinfo {title} {Cold atomic collisions: Coherent
  control of {P}enning and associative ionization},\ }\href@noop {} {\bibfield
  {journal} {\bibinfo  {journal} {Phys. Rev. Lett.}\ }\textbf {\bibinfo
  {volume} {97}},\ \bibinfo {pages} {193202} (\bibinfo {year}
  {2006})}\BibitemShut {NoStop}%
\bibitem [{\citenamefont {Gould}\ \emph {et~al.}(1988)\citenamefont {Gould},
  \citenamefont {Lett}, \citenamefont {Julienne}, \citenamefont {Phillips},
  \citenamefont {Thorsheim},\ and\ \citenamefont {Weiner}}]{Gould1988}%
  \BibitemOpen
  \bibfield  {author} {\bibinfo {author} {\bibfnamefont {P.~L.}\ \bibnamefont
  {Gould}}, \bibinfo {author} {\bibfnamefont {P.~D.}\ \bibnamefont {Lett}},
  \bibinfo {author} {\bibfnamefont {P.~S.}\ \bibnamefont {Julienne}}, \bibinfo
  {author} {\bibfnamefont {W.~D.}\ \bibnamefont {Phillips}}, \bibinfo {author}
  {\bibfnamefont {H.~R.}\ \bibnamefont {Thorsheim}},\ and\ \bibinfo {author}
  {\bibfnamefont {J.}~\bibnamefont {Weiner}},\ }\bibfield  {title} {\bibinfo
  {title} {Observation of associative ionization of ultracold laser-trapped
  sodium atoms},\ }\href@noop {} {\bibfield  {journal} {\bibinfo  {journal}
  {Phys. Rev. Lett.}\ }\textbf {\bibinfo {volume} {60}},\ \bibinfo {pages}
  {788} (\bibinfo {year} {1988})}\BibitemShut {NoStop}%
\bibitem [{\citenamefont {Oppenheimer}\ and\ \citenamefont
  {Dalgarno}(1977)}]{Oppenheimer1977}%
  \BibitemOpen
  \bibfield  {author} {\bibinfo {author} {\bibfnamefont {M.}~\bibnamefont
  {Oppenheimer}}\ and\ \bibinfo {author} {\bibfnamefont {A.}~\bibnamefont
  {Dalgarno}},\ }\bibfield  {title} {\bibinfo {title} {Associative ionization
  and interstellar {TiO$^+$} and {TiO}},\ }\href@noop {} {\bibfield  {journal}
  {\bibinfo  {journal} {Ap. J.}\ }\textbf {\bibinfo {volume} {212}},\ \bibinfo
  {pages} {683} (\bibinfo {year} {1977})}\BibitemShut {NoStop}%
\bibitem [{\citenamefont {Petrie}\ and\ \citenamefont
  {Bohme}(1994)}]{Petrie1994}%
  \BibitemOpen
  \bibfield  {author} {\bibinfo {author} {\bibfnamefont {S.}~\bibnamefont
  {Petrie}}\ and\ \bibinfo {author} {\bibfnamefont {D.~K.}\ \bibnamefont
  {Bohme}},\ }\bibfield  {title} {\bibinfo {title} {Associative ionization
  processes within interstellar clouds},\ }\href@noop {} {\bibfield  {journal}
  {\bibinfo  {journal} {Ap. J.}\ }\textbf {\bibinfo {volume} {436}},\ \bibinfo
  {pages} {411} (\bibinfo {year} {1994})}\BibitemShut {NoStop}%
\bibitem [{\citenamefont {Schofield}(2006)}]{schofield2006}%
  \BibitemOpen
  \bibfield  {author} {\bibinfo {author} {\bibfnamefont {K.}~\bibnamefont
  {Schofield}},\ }\bibfield  {title} {\bibinfo {title} {An overlooked series of
  gas phase diatomic metal oxide ions that are long-lived},\ }\href@noop {}
  {\bibfield  {journal} {\bibinfo  {journal} {J. Phys. Chem A}\ }\textbf
  {\bibinfo {volume} {110}},\ \bibinfo {pages} {6938} (\bibinfo {year}
  {2006})}\BibitemShut {NoStop}%
\bibitem [{\citenamefont {Deiß}\ \emph {et~al.}(2024)\citenamefont {Deiß},
  \citenamefont {Willitsch},\ and\ \citenamefont {Denschlag}}]{Deiß2024}%
  \BibitemOpen
  \bibfield  {author} {\bibinfo {author} {\bibfnamefont {M.}~\bibnamefont
  {Deiß}}, \bibinfo {author} {\bibfnamefont {S.}~\bibnamefont {Willitsch}},\
  and\ \bibinfo {author} {\bibfnamefont {J.~H.}\ \bibnamefont {Denschlag}},\
  }\bibfield  {title} {\bibinfo {title} {Cold trapped molecular ions and hybrid
  platforms for ions and neutral particles},\ }\href@noop {} {\bibfield
  {journal} {\bibinfo  {journal} {Nat. Phys.}\ }\textbf {\bibinfo {volume}
  {20}},\ \bibinfo {pages} {713} (\bibinfo {year} {2024})}\BibitemShut
  {NoStop}%
\bibitem [{\citenamefont {DeMille}\ \emph {et~al.}(2024)\citenamefont
  {DeMille}, \citenamefont {Hutzler}, \citenamefont {Rey},\ and\ \citenamefont
  {Zelevinsky}}]{DeMille2024}%
  \BibitemOpen
  \bibfield  {author} {\bibinfo {author} {\bibfnamefont {D.}~\bibnamefont
  {DeMille}}, \bibinfo {author} {\bibfnamefont {N.~R.}\ \bibnamefont
  {Hutzler}}, \bibinfo {author} {\bibfnamefont {A.~M.}\ \bibnamefont {Rey}},\
  and\ \bibinfo {author} {\bibfnamefont {T.}~\bibnamefont {Zelevinsky}},\
  }\bibfield  {title} {\bibinfo {title} {Quantum sensing and metrology for
  fundamental physics with molecules},\ }\href@noop {} {\bibfield  {journal}
  {\bibinfo  {journal} {Nat. Phys.}\ }\textbf {\bibinfo {volume} {20}},\
  \bibinfo {pages} {741} (\bibinfo {year} {2024})}\BibitemShut {NoStop}%
\bibitem [{\citenamefont {Loh}\ \emph {et~al.}(2011)\citenamefont {Loh},
  \citenamefont {Wang}, \citenamefont {Grau}, \citenamefont {Yahn},
  \citenamefont {Field}, \citenamefont {Greene},\ and\ \citenamefont
  {Cornell}}]{Loh2011}%
  \BibitemOpen
  \bibfield  {author} {\bibinfo {author} {\bibfnamefont {H.}~\bibnamefont
  {Loh}}, \bibinfo {author} {\bibfnamefont {J.}~\bibnamefont {Wang}}, \bibinfo
  {author} {\bibfnamefont {M.}~\bibnamefont {Grau}}, \bibinfo {author}
  {\bibfnamefont {T.~S.}\ \bibnamefont {Yahn}}, \bibinfo {author}
  {\bibfnamefont {R.~W.}\ \bibnamefont {Field}}, \bibinfo {author}
  {\bibfnamefont {C.~H.}\ \bibnamefont {Greene}},\ and\ \bibinfo {author}
  {\bibfnamefont {E.~A.}\ \bibnamefont {Cornell}},\ }\bibfield  {title}
  {\bibinfo {title} {Laser-induced fluorescence studies of {HfF$^+$} produced
  by autoionization},\ }\href@noop {} {\bibfield  {journal} {\bibinfo
  {journal} {J. Chem. Phys.}\ }\textbf {\bibinfo {volume} {135}},\ \bibinfo
  {pages} {154308} (\bibinfo {year} {2011})}\BibitemShut {NoStop}%
\bibitem [{\citenamefont {Zhang}\ \emph {et~al.}(2023)\citenamefont {Zhang},
  \citenamefont {Zhang}, \citenamefont {Bai}, \citenamefont {Ao}, \citenamefont
  {Peng}, \citenamefont {He},\ and\ \citenamefont {Tong}}]{Zhang2023}%
  \BibitemOpen
  \bibfield  {author} {\bibinfo {author} {\bibfnamefont {Y.}~\bibnamefont
  {Zhang}}, \bibinfo {author} {\bibfnamefont {Q.-Y.}\ \bibnamefont {Zhang}},
  \bibinfo {author} {\bibfnamefont {W.-L.}\ \bibnamefont {Bai}}, \bibinfo
  {author} {\bibfnamefont {Z.-Y.}\ \bibnamefont {Ao}}, \bibinfo {author}
  {\bibfnamefont {W.-C.}\ \bibnamefont {Peng}}, \bibinfo {author}
  {\bibfnamefont {S.-G.}\ \bibnamefont {He}},\ and\ \bibinfo {author}
  {\bibfnamefont {X.}~\bibnamefont {Tong}},\ }\bibfield  {title} {\bibinfo
  {title} {Generation of rotational-ground-state {HD$^+$} ions in an ion trap
  using a resonance-enhanced threshold photoionization process},\ }\href@noop
  {} {\bibfield  {journal} {\bibinfo  {journal} {Phys. Rev. A}\ }\textbf
  {\bibinfo {volume} {107}},\ \bibinfo {pages} {043101} (\bibinfo {year}
  {2023})}\BibitemShut {NoStop}%
\bibitem [{\citenamefont {Barker}\ \emph {et~al.}(2012)\citenamefont {Barker},
  \citenamefont {Antonov}, \citenamefont {Heaven},\ and\ \citenamefont
  {Peterson}}]{Barker2012}%
  \BibitemOpen
  \bibfield  {author} {\bibinfo {author} {\bibfnamefont {B.~J.}\ \bibnamefont
  {Barker}}, \bibinfo {author} {\bibfnamefont {I.~O.}\ \bibnamefont {Antonov}},
  \bibinfo {author} {\bibfnamefont {M.~C.}\ \bibnamefont {Heaven}},\ and\
  \bibinfo {author} {\bibfnamefont {K.~A.}\ \bibnamefont {Peterson}},\
  }\bibfield  {title} {\bibinfo {title} {Spectroscopic investigations of {ThF}
  and {ThF$^+$}},\ }\href@noop {} {\bibfield  {journal} {\bibinfo  {journal}
  {J. Chem. Phys.}\ }\textbf {\bibinfo {volume} {136}},\ \bibinfo {pages}
  {104305} (\bibinfo {year} {2012})}\BibitemShut {NoStop}%
\bibitem [{\citenamefont {Lau}\ \emph {et~al.}(1989)\citenamefont {Lau},
  \citenamefont {Brittain},\ and\ \citenamefont {Hildenbrand}}]{Lau1989}%
  \BibitemOpen
  \bibfield  {author} {\bibinfo {author} {\bibfnamefont {K.~H.}\ \bibnamefont
  {Lau}}, \bibinfo {author} {\bibfnamefont {R.}~\bibnamefont {Brittain}},\ and\
  \bibinfo {author} {\bibfnamefont {D.~L.}\ \bibnamefont {Hildenbrand}},\
  }\bibfield  {title} {\bibinfo {title} {High temperature thermodynamic studies
  of some gaseous thorium fluorides},\ }\href@noop {} {\bibfield  {journal}
  {\bibinfo  {journal} {J. Chem. Phys.}\ }\textbf {\bibinfo {volume} {90}},\
  \bibinfo {pages} {1158} (\bibinfo {year} {1989})}\BibitemShut {NoStop}%
\bibitem [{\citenamefont {Zhou}\ \emph {et~al.}(2019)\citenamefont {Zhou},
  \citenamefont {Ng}, \citenamefont {Cheng}, \citenamefont {Gresh},
  \citenamefont {Field}, \citenamefont {Ye},\ and\ \citenamefont
  {Cornell}}]{Zhou2019}%
  \BibitemOpen
  \bibfield  {author} {\bibinfo {author} {\bibfnamefont {Y.}~\bibnamefont
  {Zhou}}, \bibinfo {author} {\bibfnamefont {K.~B.}\ \bibnamefont {Ng}},
  \bibinfo {author} {\bibfnamefont {L.}~\bibnamefont {Cheng}}, \bibinfo
  {author} {\bibfnamefont {D.~N.}\ \bibnamefont {Gresh}}, \bibinfo {author}
  {\bibfnamefont {R.~W.}\ \bibnamefont {Field}}, \bibinfo {author}
  {\bibfnamefont {J.}~\bibnamefont {Ye}},\ and\ \bibinfo {author}
  {\bibfnamefont {E.~A.}\ \bibnamefont {Cornell}},\ }\bibfield  {title}
  {\bibinfo {title} {Visible and ultraviolet laser spectroscopy of {ThF}},\
  }\href@noop {} {\bibfield  {journal} {\bibinfo  {journal} {J. Mol.
  Spectrosc.}\ }\textbf {\bibinfo {volume} {358}},\ \bibinfo {pages} {1}
  (\bibinfo {year} {2019})}\BibitemShut {NoStop}%
\bibitem [{\citenamefont {Fleig}(2017)}]{Fleig2017}%
  \BibitemOpen
  \bibfield  {author} {\bibinfo {author} {\bibfnamefont {T.}~\bibnamefont
  {Fleig}},\ }\bibfield  {title} {\bibinfo {title} {{TaO$^+$} as a candidate
  molecular ion for searches of physics beyond the standard model},\
  }\href@noop {} {\bibfield  {journal} {\bibinfo  {journal} {Phys. Rev. A}\
  }\textbf {\bibinfo {volume} {95}},\ \bibinfo {pages} {022504} (\bibinfo
  {year} {2017})}\BibitemShut {NoStop}%
\bibitem [{\citenamefont {Luo}(2007)}]{Luo2007}%
  \BibitemOpen
  \bibfield  {author} {\bibinfo {author} {\bibfnamefont {Y.~R.}\ \bibnamefont
  {Luo}},\ }\href@noop {} {\emph {\bibinfo {title} {Comprehensive handbook of
  chemical bond energies}}}\ (\bibinfo  {publisher} {CRC Press, Boca Raton, FL,
  USA},\ \bibinfo {year} {2007})\BibitemShut {NoStop}%
\bibitem [{\citenamefont {Dyke}\ \emph {et~al.}(1987)\citenamefont {Dyke},
  \citenamefont {Ellis}, \citenamefont {Feh\'er}, \citenamefont {Morris},
  \citenamefont {Paul},\ and\ \citenamefont {Stevens}}]{Dyke1987}%
  \BibitemOpen
  \bibfield  {author} {\bibinfo {author} {\bibfnamefont {J.~M.}\ \bibnamefont
  {Dyke}}, \bibinfo {author} {\bibfnamefont {A.~M.}\ \bibnamefont {Ellis}},
  \bibinfo {author} {\bibfnamefont {M.}~\bibnamefont {Feh\'er}}, \bibinfo
  {author} {\bibfnamefont {A.}~\bibnamefont {Morris}}, \bibinfo {author}
  {\bibfnamefont {A.~J.}\ \bibnamefont {Paul}},\ and\ \bibinfo {author}
  {\bibfnamefont {J.~C.~H.}\ \bibnamefont {Stevens}},\ }\bibfield  {title}
  {\bibinfo {title} {High-temperature photoelectron spectroscopy; a study of
  niobium monoxide and tantalum monoxide},\ }\href@noop {} {\bibfield
  {journal} {\bibinfo  {journal} {J. Chem. Soc., Faraday Trans.}\ }\textbf
  {\bibinfo {volume} {83}},\ \bibinfo {pages} {1555} (\bibinfo {year}
  {1987})}\BibitemShut {NoStop}%
\bibitem [{\citenamefont {Sorensen}\ \emph {et~al.}(2020)\citenamefont
  {Sorensen}, \citenamefont {Tieu}, \citenamefont {Sevy}, \citenamefont
  {Merriles}, \citenamefont {Nielson}, \citenamefont {Ewigleben},\ and\
  \citenamefont {Morse}}]{Sorensen2020}%
  \BibitemOpen
  \bibfield  {author} {\bibinfo {author} {\bibfnamefont {J.~J.}\ \bibnamefont
  {Sorensen}}, \bibinfo {author} {\bibfnamefont {E.}~\bibnamefont {Tieu}},
  \bibinfo {author} {\bibfnamefont {A.}~\bibnamefont {Sevy}}, \bibinfo {author}
  {\bibfnamefont {D.~M.}\ \bibnamefont {Merriles}}, \bibinfo {author}
  {\bibfnamefont {C.}~\bibnamefont {Nielson}}, \bibinfo {author} {\bibfnamefont
  {J.~C.}\ \bibnamefont {Ewigleben}},\ and\ \bibinfo {author} {\bibfnamefont
  {M.~D.}\ \bibnamefont {Morse}},\ }\bibfield  {title} {\bibinfo {title} {Bond
  dissociation energies of transition metal oxides: {CrO}, {MoO}, {RuO}, and
  {RhO}},\ }\href@noop {} {\bibfield  {journal} {\bibinfo  {journal} {J. Chem.
  Phys.}\ }\textbf {\bibinfo {volume} {153}},\ \bibinfo {pages} {074303}
  (\bibinfo {year} {2020})}\BibitemShut {NoStop}%
\bibitem [{\citenamefont {Ghiassee}\ \emph {et~al.}(2023)\citenamefont
  {Ghiassee}, \citenamefont {Christensen}, \citenamefont {Fenn},\ and\
  \citenamefont {Armentrout}}]{Ghiassee2023}%
  \BibitemOpen
  \bibfield  {author} {\bibinfo {author} {\bibfnamefont {M.}~\bibnamefont
  {Ghiassee}}, \bibinfo {author} {\bibfnamefont {E.~G.}\ \bibnamefont
  {Christensen}}, \bibinfo {author} {\bibfnamefont {T.}~\bibnamefont {Fenn}},\
  and\ \bibinfo {author} {\bibfnamefont {P.~B.}\ \bibnamefont {Armentrout}},\
  }\bibfield  {title} {\bibinfo {title} {Guided ion beam studies of the
  {Dy}+{O} $\rightarrow$ {DyO$^+$} + {e$^-$} chemi-ionization reaction
  thermochemistry and dysprosium oxide, carbide, sulfide, dioxide, and
  sulfoxide cation bond energies},\ }\href@noop {} {\bibfield  {journal}
  {\bibinfo  {journal} {J. Phys. Chem A}\ }\textbf {\bibinfo {volume} {127}},\
  \bibinfo {pages} {169} (\bibinfo {year} {2023})}\BibitemShut {NoStop}%
\bibitem [{\citenamefont {Popa}\ \emph {et~al.}(2024)\citenamefont {Popa},
  \citenamefont {Schaller}, \citenamefont {Fielicke}, \citenamefont {Lim},
  \citenamefont {Sartakov}, \citenamefont {Tarbutt},\ and\ \citenamefont
  {Meijer}}]{Popa2024}%
  \BibitemOpen
  \bibfield  {author} {\bibinfo {author} {\bibfnamefont {S.}~\bibnamefont
  {Popa}}, \bibinfo {author} {\bibfnamefont {S.}~\bibnamefont {Schaller}},
  \bibinfo {author} {\bibfnamefont {A.}~\bibnamefont {Fielicke}}, \bibinfo
  {author} {\bibfnamefont {J.}~\bibnamefont {Lim}}, \bibinfo {author}
  {\bibfnamefont {B.~G.}\ \bibnamefont {Sartakov}}, \bibinfo {author}
  {\bibfnamefont {M.~R.}\ \bibnamefont {Tarbutt}},\ and\ \bibinfo {author}
  {\bibfnamefont {G.}~\bibnamefont {Meijer}},\ }\bibfield  {title} {\bibinfo
  {title} {Understanding inner-shell excitations in molecules through
  spectroscopy of the 4{\it f} hole states of {Y}b{F}},\ }\href@noop {}
  {\bibfield  {journal} {\bibinfo  {journal} {Phys. Rev. X}\ }\textbf {\bibinfo
  {volume} {14}},\ \bibinfo {pages} {021035} (\bibinfo {year}
  {2024})}\BibitemShut {NoStop}%
\bibitem [{\citenamefont {Linton}\ \emph {et~al.}(1986)\citenamefont {Linton},
  \citenamefont {Gaudet},\ and\ \citenamefont {Schall}}]{Linton1986}%
  \BibitemOpen
  \bibfield  {author} {\bibinfo {author} {\bibfnamefont {C.}~\bibnamefont
  {Linton}}, \bibinfo {author} {\bibfnamefont {D.~M.}\ \bibnamefont {Gaudet}},\
  and\ \bibinfo {author} {\bibfnamefont {H.}~\bibnamefont {Schall}},\
  }\bibfield  {title} {\bibinfo {title} {Laser spectroscopy of dysprosium
  monoxide: observation and analysis of several low-lying electronic states},\
  }\href@noop {} {\bibfield  {journal} {\bibinfo  {journal} {J. Mol.
  Spectrosc.}\ }\textbf {\bibinfo {volume} {115}},\ \bibinfo {pages} {58}
  (\bibinfo {year} {1986})}\BibitemShut {NoStop}%
\bibitem [{\citenamefont {Loh}\ \emph {et~al.}(2012)\citenamefont {Loh},
  \citenamefont {Stutz}, \citenamefont {Yahn}, \citenamefont {Looser},
  \citenamefont {Field},\ and\ \citenamefont {Cornell}}]{Loh2012}%
  \BibitemOpen
  \bibfield  {author} {\bibinfo {author} {\bibfnamefont {H.}~\bibnamefont
  {Loh}}, \bibinfo {author} {\bibfnamefont {R.~P.}\ \bibnamefont {Stutz}},
  \bibinfo {author} {\bibfnamefont {T.~S.}\ \bibnamefont {Yahn}}, \bibinfo
  {author} {\bibfnamefont {H.}~\bibnamefont {Looser}}, \bibinfo {author}
  {\bibfnamefont {R.~W.}\ \bibnamefont {Field}},\ and\ \bibinfo {author}
  {\bibfnamefont {E.~A.}\ \bibnamefont {Cornell}},\ }\bibfield  {title}
  {\bibinfo {title} {{REMPI} spectroscopy of {HfF}},\ }\href@noop {} {\bibfield
   {journal} {\bibinfo  {journal} {J. Mol. Spectrosc.}\ }\textbf {\bibinfo
  {volume} {276-277}},\ \bibinfo {pages} {49} (\bibinfo {year}
  {2012})}\BibitemShut {NoStop}%
\bibitem [{\citenamefont {Cheng}(1992)}]{Cheng1992}%
  \BibitemOpen
  \bibfield  {author} {\bibinfo {author} {\bibfnamefont {C.-H.}\ \bibnamefont
  {Cheng}},\ }\emph {\bibinfo {title} {Laser spectroscopic studies of holmium
  monoxide and dysprosium monoxide}},\ \href@noop {} {Ph.D. thesis},\ \bibinfo
  {school} {The University of New Brunswick} (\bibinfo {year}
  {1992})\BibitemShut {NoStop}%
\bibitem [{\citenamefont {Merkt}(1997)}]{Merkt1997}%
  \BibitemOpen
  \bibfield  {author} {\bibinfo {author} {\bibfnamefont {F.}~\bibnamefont
  {Merkt}},\ }\bibfield  {title} {\bibinfo {title} {Molecules in high {R}ydberg
  states},\ }\href@noop {} {\bibfield  {journal} {\bibinfo  {journal} {Annu.
  Rev. Phys. Chem.}\ }\textbf {\bibinfo {volume} {48}},\ \bibinfo {pages} {675}
  (\bibinfo {year} {1997})}\BibitemShut {NoStop}%
\bibitem [{\citenamefont {Zhu}\ and\ \citenamefont {Johnson}(1991)}]{Zhu1991}%
  \BibitemOpen
  \bibfield  {author} {\bibinfo {author} {\bibfnamefont {L.}~\bibnamefont
  {Zhu}}\ and\ \bibinfo {author} {\bibfnamefont {P.}~\bibnamefont {Johnson}},\
  }\bibfield  {title} {\bibinfo {title} {Mass analyzed threshold ionization
  spectroscopy},\ }\href@noop {} {\bibfield  {journal} {\bibinfo  {journal} {J.
  Chem. Phys.}\ }\textbf {\bibinfo {volume} {94}},\ \bibinfo {pages} {5769}
  (\bibinfo {year} {1991})}\BibitemShut {NoStop}%
\bibitem [{\citenamefont {Ackermann}\ \emph {et~al.}(1976)\citenamefont
  {Ackermann}, \citenamefont {Rauh},\ and\ \citenamefont
  {Thorn}}]{Ackermann1976}%
  \BibitemOpen
  \bibfield  {author} {\bibinfo {author} {\bibfnamefont {R.~J.}\ \bibnamefont
  {Ackermann}}, \bibinfo {author} {\bibfnamefont {E.~G.}\ \bibnamefont
  {Rauh}},\ and\ \bibinfo {author} {\bibfnamefont {R.~J.}\ \bibnamefont
  {Thorn}},\ }\bibfield  {title} {\bibinfo {title} {The thermodynamics of
  ionization of gaseous oxides; the first ionization potentials of the
  lanthanide metals and monoxides},\ }\href@noop {} {\bibfield  {journal}
  {\bibinfo  {journal} {J. Chem. Phys.}\ }\textbf {\bibinfo {volume} {65}},\
  \bibinfo {pages} {1027} (\bibinfo {year} {1976})}\BibitemShut {NoStop}%
\bibitem [{\citenamefont {Studer}\ \emph {et~al.}(2017)\citenamefont {Studer},
  \citenamefont {Dyrauf}, \citenamefont {Naubereit}, \citenamefont {Heinke},\
  and\ \citenamefont {Wendt}}]{Studer2017}%
  \BibitemOpen
  \bibfield  {author} {\bibinfo {author} {\bibfnamefont {D.}~\bibnamefont
  {Studer}}, \bibinfo {author} {\bibfnamefont {P.}~\bibnamefont {Dyrauf}},
  \bibinfo {author} {\bibfnamefont {P.}~\bibnamefont {Naubereit}}, \bibinfo
  {author} {\bibfnamefont {R.}~\bibnamefont {Heinke}},\ and\ \bibinfo {author}
  {\bibfnamefont {K.}~\bibnamefont {Wendt}},\ }\bibfield  {title} {\bibinfo
  {title} {Resonance ionization spectroscopy in dysprosium; excitation scheme
  development and re-determination of the first ionization potential},\
  }\href@noop {} {\bibfield  {journal} {\bibinfo  {journal} {Hyperfine
  Interact.}\ }\textbf {\bibinfo {volume} {238}},\ \bibinfo {pages} {8}
  (\bibinfo {year} {2017})}\BibitemShut {NoStop}%
\end{thebibliography}

\end{document}